\documentclass[prl,superscriptaddress,citeautoscript,twocolumn]{revtex4-1}

\usepackage{latexsym}
\usepackage{graphics}
\usepackage{graphicx}
\usepackage{amsmath,amssymb}
\usepackage{color}
\usepackage{empheq}
\usepackage{tcolorbox}
\usepackage{wrapfig}
\usepackage{blindtext}
\usepackage{float}
\usepackage[hidelinks,colorlinks=true,linkcolor=blue,urlcolor=blue,citecolor=blue,anchorcolor=blue]{hyperref}

\usepackage{bibunits}

\defaultbibliography{ssb.bib}
\defaultbibliographystyle{apsrev4-1}

\DeclareMathOperator{\sgn}{sgn}

\begin{document}
\begin{bibunit}
 
\title{Spontaneous symmetry breaking in a coherently driven nanophotonic Bose-Hubbard dimer}
\author{B. Garbin}
\affiliation{Universit\'e Paris-Saclay, CNRS, Centre de Nanosciences et de Nanotechnologies, 91120 Palaiseau, France}
\author{A. Giraldo}
\affiliation{Department of Mathematics and Dodd-Walls Centre, The University of Auckland, Private Bag 92019, Auckland 1142, New Zealand}
\author{K. J. H. Peters}
\affiliation{Center  for  Nanophotonics,  AMOLF,  Science  Park  104,  1098  XG  Amsterdam,  The  Netherlands}
\author{N. G. R. Broderick}
\affiliation{Department of Physics and Dodd-Walls Centre, The University of Auckland, Private Bag 92019, Auckland 1142, New Zealand}
\affiliation{Photon Factory, Department of Physics, University of Auckland, Auckland 1010, New Zealand}
\author{A. Spakman}
\affiliation{Center  for  Nanophotonics,  AMOLF,  Science  Park  104,  1098  XG  Amsterdam,  The  Netherlands}
\author{F. Raineri}
\affiliation{Universit\'e Paris-Saclay, CNRS, Centre de Nanosciences et de Nanotechnologies, 91120 Palaiseau, France}
\affiliation{Universit\'e de Paris, Physics department, Paris (France)}
\author{A. Levenson}
\affiliation{Universit\'e Paris-Saclay, CNRS, Centre de Nanosciences et de Nanotechnologies, 91120 Palaiseau, France}
\author{S. R. K.  Rodriguez}
\affiliation{Center  for  Nanophotonics,  AMOLF,  Science  Park  104,  1098  XG  Amsterdam,  The  Netherlands}
\author{B. Krauskopf}
\affiliation{Department of Mathematics and Dodd-Walls Centre, The University of Auckland, Private Bag 92019, Auckland 1142, New Zealand}
\author{A. M. Yacomotti}
\affiliation{Universit\'e Paris-Saclay, CNRS, Centre de Nanosciences et de Nanotechnologies, 91120 Palaiseau, France}

\begin{abstract}
We report on the first experimental observation of spontaneous mirror symmetry breaking (SSB) in coherently driven-dissipative coupled optical cavities. SSB is observed as the breaking of the spatial or mirror $\mathbb{Z}_2$ symmetry between two symmetrically pumped and evanescently coupled photonic crystal nanocavities, and manifests itself as random intensity localization in one of the two cavities. We show that, in a system featuring repulsive boson interactions ($U>0$), the observation of a pure pitchfork bifurcation requires negative photon hopping energies ($J<0$), which we have realized in our photonic crystal molecule. SSB is observed over a wide range of the two-dimensional parameter space of driving intensity and detuning, where we also find a region that exhibits bistable symmetric behavior. Our results pave the way for the experimental study of limit cycles and deterministic chaos arising from SSB, as well as the study of nonclassical photon correlations close to SSB transitions.
\end{abstract}

\date{Received: \today}

\maketitle


The Bose-Hubbard dimer (BHD) is a paradigmatic system featuring quantum dynamics of bosons hopping across sites and interacting on-site~\cite{fisher_boson_1989}. The competition between these processes in the \textit{closed} BHD unveiled a wealth of quantum and classical nonlinear phenomena, including self-trapping~\cite{albiez_direct_2005} and symmetry breaking~\cite{zibold_classical_2010}. In most real BHDs, however, particle losses need to be compensated by external driving~\cite{carusotto_quantum_2013,schmidt_circuit_2013,hartmann_quantum_2016,noh_quantum_2016}. These \textit{driven-dissipative} BHDs have recently drawn much interest, as they display intriguing classical~\cite{coullet_chaotic_2001,giraldo_the_2020} and quantum~\cite{minganti_spectral_2018} phenomena due to the balance between driving and dissipation. Driven-dissipative BHDs have been implemented on light-matter systems like semiconductor microcavities~\cite{lagoudakis_coherent_2010,abbarchi_macroscopic_2013,rodriguez_interaction-induced_2016} and superconducting circuits~\cite{raftery_observation_2014,eichler_quantum-limited_2014}. 

SSB is a universal phenomenon occurring when a symmetric system ends up in an asymmetric state. SSB and its applications have attracted significant interest in optics~\cite{kaplan_enhancement_1981, malomed_spontaneous_2013, liu_spontaneous_2014,  hamel_spontaneous_2015,cao_experimental_2017, delbino_symmetry_2017,xu_spontaneous_2021}.  In single coherently driven optical cavities, SSB emerges from the nonlinear coupling between polarization or counter-propagating modes~\cite{garbin_asymmetric_2020, delbino_symmetry_2017}. SSB has also been observed in coupled active photonic crystal nano-resonators~\cite{hamel_spontaneous_2015},  which share some properties with incoherently driven BHDs~\cite{secli_signatures_2021}.  However, optical gain saturation---a nonlinear mechanism reminiscent of inelastic two body collisions in atomic systems~\cite{coullet_chaotic_2001}---results in a fundamental difference with respect to BHDs. Compared to the incoherent pumping case, coherent driving offers a more versatile parameter control and suppressed spontaneous emission noise. Furthermore, coherently driven BHD have been predicted to undergo entanglement close to the SSB transition~\cite{casteels_quantum_2017}, thereby bridging classical nonlinear dynamics and quantum optics. 

Despite the strong interest in coherently driven dissipative BHD physics, SSB remains unreported. This is likely due to two reasons.  First, conventional driven-dissipative BHDs display symmetry broken phases restricted to narrow parameter regions, which coexist with other optical bistabilities~\cite{cao_two_2016,giraldo_the_2020}. Thus, accessing SSB is challenging. Second, resonant  optical  excitation  experiments require mode-matching, thereby leading to more stringent conditions than their incoherent pumping counterparts. 

Here we demonstrate coherently driven SSB in a photonic crystal driven-dissipative  BHD. Unlike standard BHDs, our system features a negative photon hopping energy that, in conjunction with a positive on-site interaction energy, results in a broken symmetry phase across large parameter regions~\cite{giraldo_the_2020}. These are {\em pure} pitchfork-bifurcated phases; they do not coexist with a bistable homogeneous state which might otherwise hinder the SSB observation.

\begin{figure}[t]
  \includegraphics[width=1 \columnwidth]{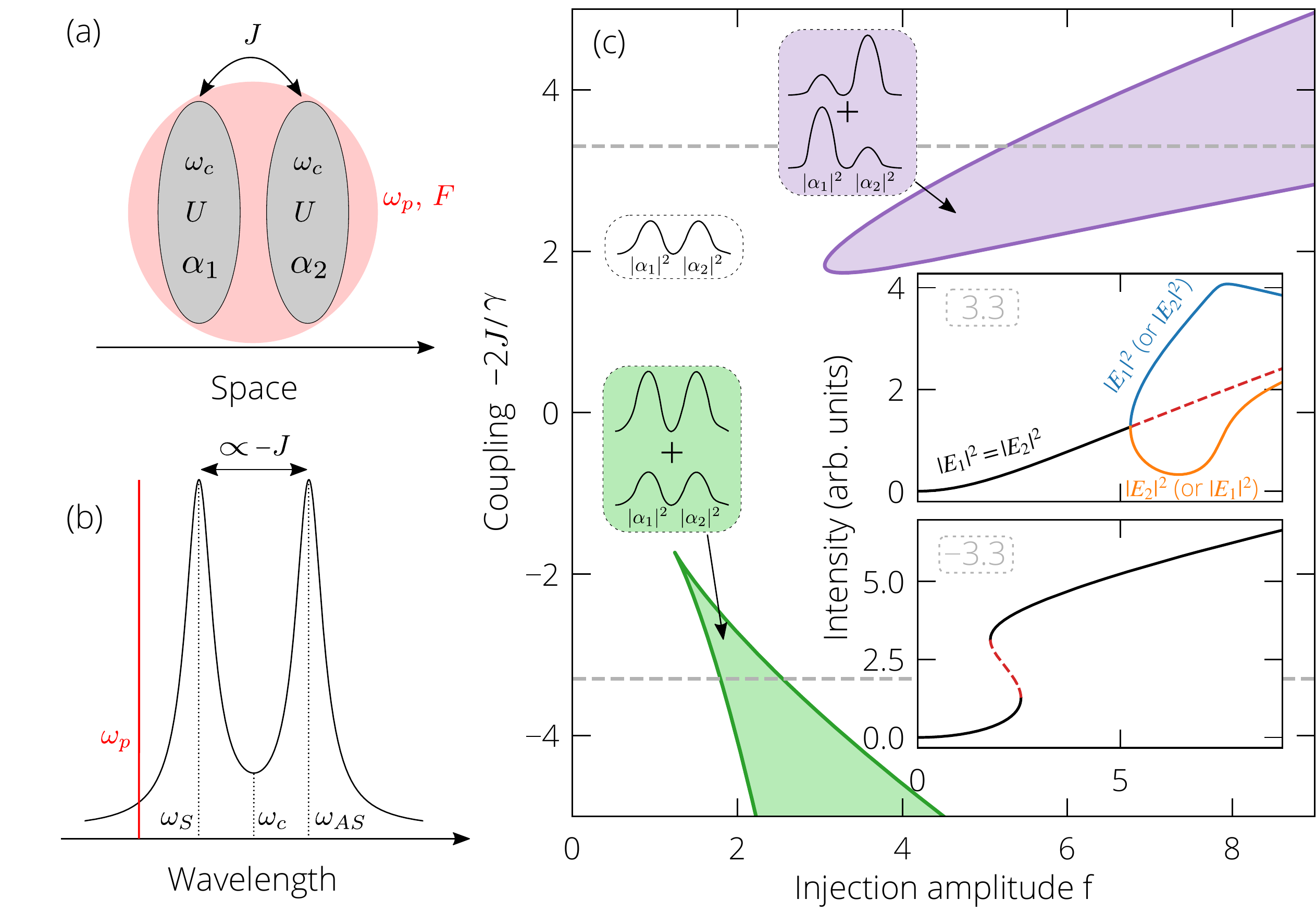}
  \caption{Illustration of (a) two evanescently-coupled injected resonators and (b) the resulting split hybrid symmetric ($\omega_S$) and anti-symmetric ($\omega_{AS}$) modes. (c) Two-parameter bifurcation diagram for $\Delta=0$ showing the regions of bistability of symmetric (green region) and asymmetric (purple region) steady states. The insets represent bifurcation slices showcasing the existence of pitchfork (top, $-2J/\gamma = +3.3$) and saddle-node (bottom, $-2J/\gamma = -3.3$) bifurcations.}
  \label{fig1}
\end{figure}

We consider a BHD of two sites---``or cavities''---where photons interact on-site with energy $U>0$, hop across sites with energy $J<0$, and that is coherently driven symmetrically with amplitude $F$ and frequency $\omega_p$ [see Fig.~\ref{fig1}(a)]. We apply the truncated Wigner approximation~\cite{carusotto_quantum_2013}, neglecting quantum correlations but accounting for quantum fluctuations via stochastic terms in semiclassical equations of motion.  In a frame rotating at $\omega_p$, the expectation values $\alpha_{1,2}$ of the bosonic operators $\hat a_1$ and $\hat a_2$ in the two sites satisfy:

\begin{equation}
  \begin{split}
    i\frac{d\alpha_{1,2}}{d t} &= \left(-\Delta - i\frac{\gamma}{2}+ 2U |\alpha_{1,2}|^2\right) \alpha_{1,2} -J \alpha_{2,1}\\
    & +F_{1,2} + \sqrt{\frac{\gamma}{2}}\zeta_{1,2}(t).
  \end{split}  \label{eq:BHM}
\end{equation}
\noindent Here,  $\Delta=\omega_p-\omega_c$ is the detuning between $\omega_p$ and the cavities resonance frequency $\omega_c$, and $F_{1}=F_2=F$ are coherent driving amplitude applied to each cavity.  $\gamma$ is the loss rate. The last term in Eqs.~\eqref{eq:BHM} models the fluctuations as complex Gaussian noise terms $\zeta_j(t)=[\zeta_j'(t)+i\zeta_j''(t)]/\sqrt{2}$ ($j=1,2$) which have zero mean and are delta-correlated in $t$. 

In our experimental implementation, the sites are given by two evanescently-coupled photonic crystal (PhC) nanocavities with embedded quantum wells (QWs) illuminated by an optically resonant laser field. The bosons are intracavity photons that experience carrier-induced Kerr-like nonlinearities below the QW optical transparency, leading to a blue-shifting intensity-dependent refractive index ($U>0$). Using Eqs.~\eqref{eq:BHM} to model the PhC dimer, we assume ultrafast carrier dynamics instantaneous Kerr nonlinearity.

For identical cavities with $2|J|>\gamma$ and in the linear regime, two hybrid modes arise [Fig.~\ref{fig1}(b)]: the symmetric mode with $\alpha_1 = \alpha_2$, and the anti-symmetric mode  with $\alpha_1=-\alpha_2$; the mode splitting and their relative spectral position, respectively, depend on the magnitude and sign of $J$. Upon coherent in-phase symmetric illumination of the two cavities, only the symmetric mode is excited. The sign of $J$ has a large impact on the observed behavior [Fig.~\ref{fig1}(c)]: provided $U>0$, pure SSB only exists for $J<0$ (purple region and top inset). For $J>0$, the bistability of the homogeneous symmetric states prevails (green region and bottom inset), see Fig.~\ref{fig1}(c) for $\Delta=0$. 

Decreasing the value of $\Delta$ from zero results in an overlap of these two regions near $J=0$; this creates regions of e.g. periodic and chaotic behavior (see Supplementary Material). For $J>0$, however, the SSB transition always lies inside the bistable symmetric zone. On the contrary, the pure super-critical pitchfork bifurcation is only warranted for $J<0$, which we can implement experimentally.

\begin{figure}[t]
  \includegraphics[width=1 \columnwidth]{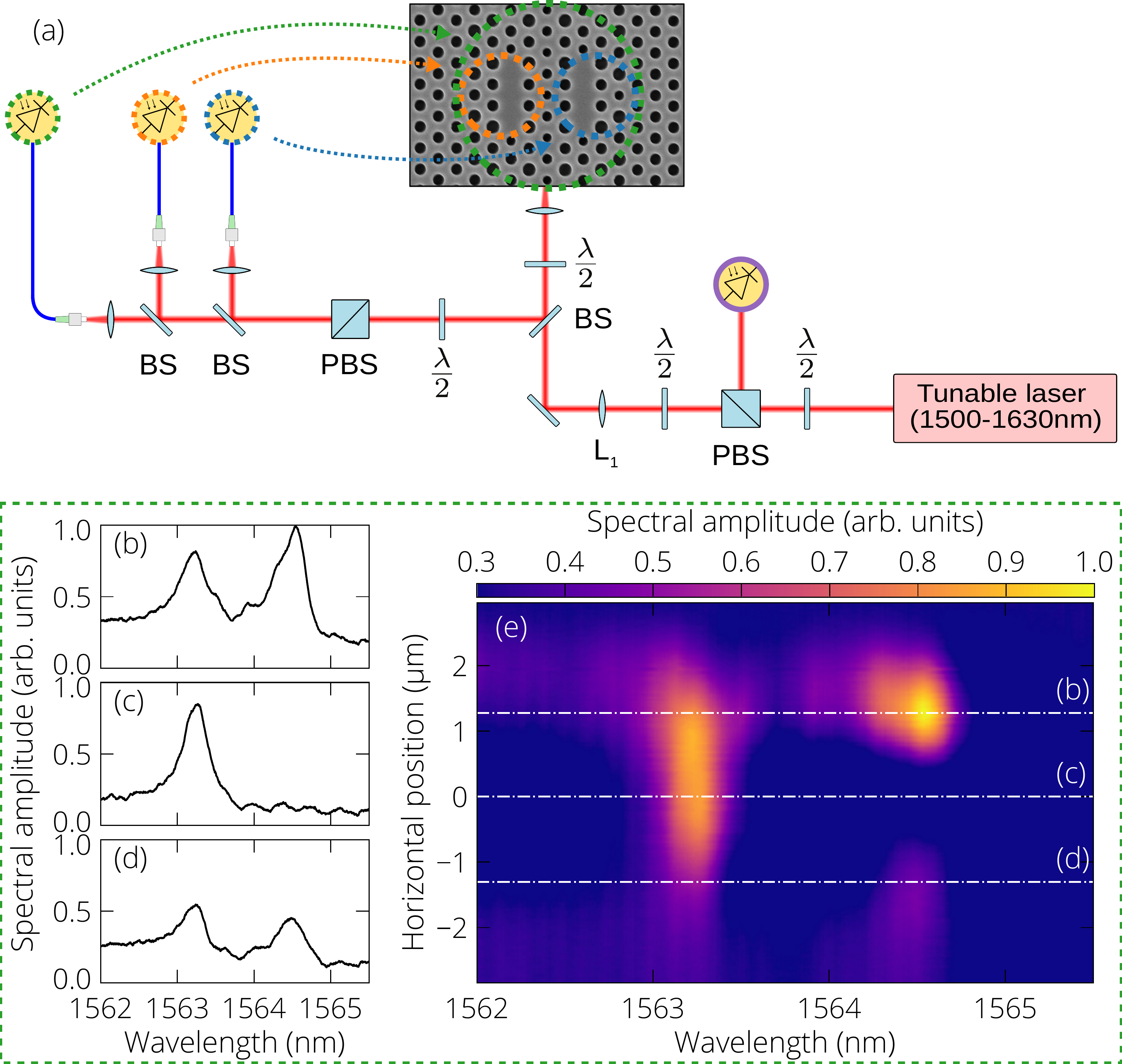}
  \caption{Experimental setup and operational regime. (a) Schematic of the experimental setup. BS, beam splitter; PBS, polarizing beam splitter; L$_1$, 100~cm focal lens; linear spectroscopy (dashed green circles) and individual cavity output (blue and orange dashed circles). (b)-(d) Selected spectra for given sample horizontal positions. (e) Evolution of the spectra with the sample horizontal position.}
  \label{fig2}
\end{figure}

Figure{~\ref{fig2}} shows our experimental setup. Two evanescently-coupled photonic-crystal nanoresonators stand on a free-standing InP membrane with four embedded InGaAs/InGaAsP QWs. The lattice period $a$ is 435~nm (hole radii $r_0 = 116$~nm). Spatial mode matching of the illumination laser beam to the cavity mode is enhanced by means of far-field engineering of the PhC cavity~\cite{haddadi_photonic_2014} (hole radii $r_2 = 137$~nm). Such mode matching is further improved using a lens with 100~cm focal length (L$_1$ in Fig.{~\ref{fig2}}). Importantly, negative $J$ with targeted splitting is achieved by controlling the radii of selected holes~\cite{hamel_spontaneous_2015} (hole radii $r_{0,2}-20\%$). The two cavities are coherently driven close to their resonant wavelength ($\approx$ 1563.9~nm) with a narrow-linewidth widely-tunable laser, whose beam is split for injection power monitoring. Finally, a $100\times$ magnification with 0.95 numerical aperture microscope objective (Olympus MPLAN x100 IR) focuses the laser beam to a 2.2~$\mu$m diameter spot on the sample from the top. The sample is mounted on a PZT driven 6-axis translation stage for controlling the relative position between the driving beam and the cavities and, hence, the relative power impinging on the two cavities. We use a half-wave plate to rotate the linearly polarized injection beam at input, by 45$^\circ$ from the cavities polarization. The cavities outputs are subsequently separated from the reflected part of the injection beam with polarization optics, and split to allow spectral and temporal analysis. The spectral part is directed to a liquid-nitrogen-cooled spectrometer (Acton SP 2500). To independently analyze three different regions of the sample, the temporal part is split and directed to: (i) a 750~Hz bandwidth Newfocus fW detector for linear spectroscopy (large dashed circle in Fig.~\ref{fig2}), and (ii) two 550~MHz bandwidth low-noise APD detectors (Princeton Lightwave PLA-8XX, N.E.P.=250 fW/$\sqrt \mathrm{Hz}$) for measuring the output intensity of each cavity (smaller dashed circles in Fig.~\ref{fig2}).

We first perform linear reflectivity spectroscopy by sweeping the driving laser wavelength. Figures{~\ref{fig2}}(b)-(d) show spectra for selected horizontal positions of the driving beam, measured with the fW detector. The relative excitation of the two hybrid modes depends on the spatial location of the driving beam relative to the sample. For an off-centered beam [panels (b) and (d)] both modes are excited. For a centered  beam [panel (c)], only the symmetric mode is excited. Figure{~\ref{fig2}}(e) shows a collection of such spectra with the driving position varied along the horizontal axes. It allows us to identify the central driving condition (horizontal position labelled 0~$\mu$m) which minimizes driving asymmetries [$F_1 \approx F_2$, see Eqs.~\eqref{eq:BHM}].

Since $J$ is negative, the symmetric mode is located at shorter wavelength (1563.2~nm) than the anti-symmetric one (1564.5~nm). Based on the measured mode splitting (1.3~nm) and mode linewidth (0.39~nm), we estimate the normalized coupling constant $\kappa=-2J/\gamma \approx 3.3$; this is compatible with SSB observation for low excitation power. Increasing the splitting to $|\kappa|\gtrsim 5$, for instance without barrier hole modification, increases the injection amplitude of the SSB threshold and might hinder observation [see Fig.~\ref{fig1}(c)].

\begin{figure}[t]
  \centerline{\includegraphics[width=1 \columnwidth]{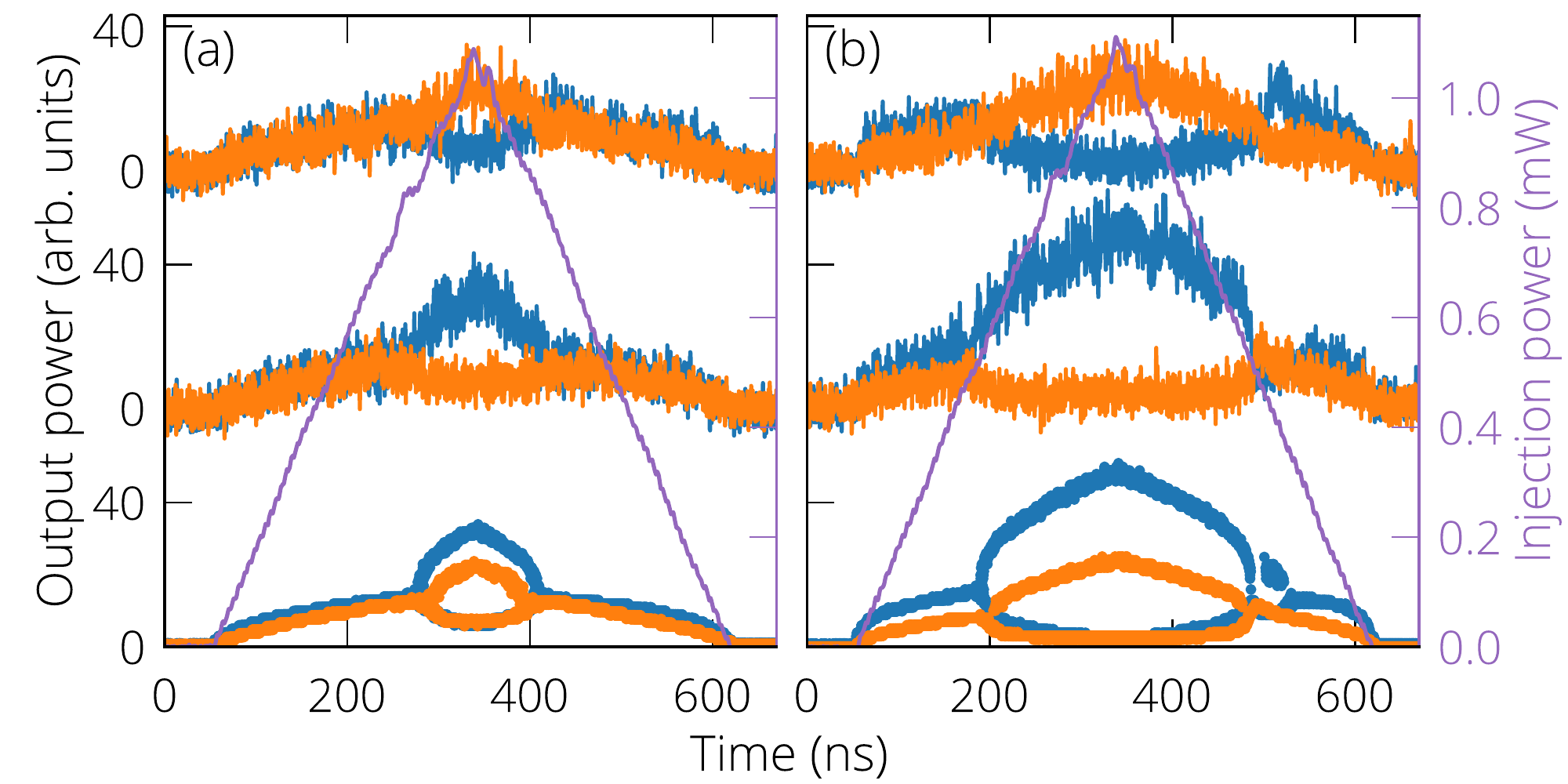}}
  \caption{Temporal observation of SSB. (a), (b) Top and middle traces are subsequent realizations of the same experiment. Bottom traces are histogram maxima computed from 500 realizations; shown are the output powers of the two cavities (blue and orange, respectively) and the driving power (purple curve). (a) $\delta=-1.59$. (b) $\delta=-2.86$.}
  \label{fig3}
\end{figure}

\begin{figure*}[t]
  \centerline{\includegraphics[width=2. \columnwidth]{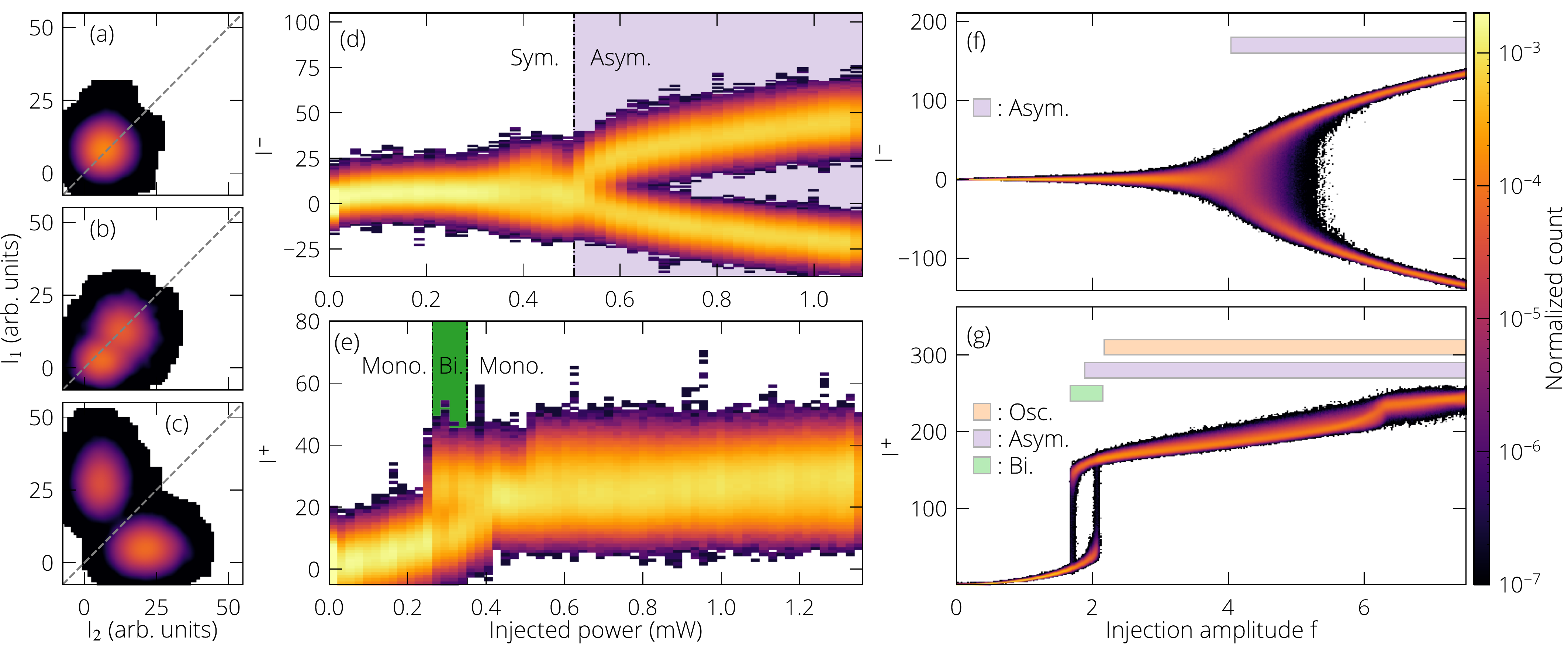}}
  \caption{Identification of operational regimes. (a)-(c) Selected projections of the phase space showing monostable symmetric, bistable symmetric, and bistable asymmetric steady states, respectively. (d), (e) Experimental and (f), (g) numerical histograms computed from 500 realizations, showing $I^-$ [(d), (f)] and $I^+$ [(e), (g)] versus injection strength. (a) 0.22~mW. (b) 0.26~mW. (c) 1.03~mW. (a), (c) $\delta=-0.82$. (d), (f) $\delta=-2.86$. (b), (e), (g) $\delta=-6.19$. }
  \label{fig4}
\end{figure*}

After locating the illumination spot at the center of the dimer [line (c) in Fig.~\ref{fig2}(e)], the excitation power (purple traces) is ramped up to about 1.5~mW within $\sim 300$~ns to minimize thermal effects. Figures{~\ref{fig3}}(a,b) show typical cavity intensity time traces (blue and orange curves) for two different values of the normalized detuning $\delta=-2\Delta/\gamma$. At low driving powers, the two cavity output traces are coincident, indicating that a delocalized (symmetric) hybrid mode is being excited. At time of about 285~ns, however, the two output intensities part, which we identify as the SSB transition point. 
The spontaneous character of the SSB manifests itself as randomly selected asymmetric states (Fig. ~\ref{fig3}, top and middle traces). The bottom traces in Fig.{~\ref{fig3}}, built by overlapping maxima of histograms of 500 successive realizations, evidence the mirror symmetry of the two asymmetric states (see Supplementary Material for comparisons with numerical simulations), and can be seen as the reconstruction of the underlying pitchfork bifurcation. 

When increasing the magnitude of the detuning from Fig.~\ref{fig3}(a) to (b), we need to realign the illumination spot to observe randomly selected asymmetric states. This may be understood from Eqs.~\eqref{eq:BHM} in the presence of asymmetries in the cavity frequencies ($\omega_{c,1} \neq \omega_{c,2}$) and driving fields ($F_1 \neq F_2$). The first one is due to unavoidable fabrication imperfections ($\approx$~0.13~$\pm$~0.07~nm), while the second one accounts for deviations of the illumination spot from the dimer center. Indeed, numerical simulations (not shown) indicate that the asymmetry parameter $\xi\sim F_1-F_2$ that induces random switching changes with $\Delta$, consistent with the experimental observation; this explains the difference in cavity intensities when moving from panel (a) to (b) related to slight misalignment in the optical detection devices. The observation of SBB through compensation of two asymmetry parameters has recently been reported in ~\cite{garbin_asymmetric_2020}.

Figure{~\ref{fig4}} illustrates the three main regimes of operation observed in our experiment, as well as the associated SSB transition and bistability. Projections of the phase space onto the $(I_1, I_2)$ plane of the two measured intensities, for given values of injection power and detuning, show: (i) equal intensity in the two cavities [Fig.{~\ref{fig4}}(a)], (ii) bistable operation with symmetric cavity intensities [Fig.{~\ref{fig4}}(b)], and (iii) bistable operation with asymmetric cavity intensities [symmetry broken states, Fig.{~\ref{fig4}}(c)]. To identify the different phases and transitions as parameters change, we consider the intensity difference $I^- = I_1 - I_2$ and sum $I^+ = I_1 + I_2$, which indicate the presence of asymmetric and symmetric states by highlighting anticorrelations and correlations, respectively (see Supplementary Material). To identify the number of states as the injected power is varied, we analyze, for two given values of detuning, the number of maxima in $I^-$ and $I^+$ [Fig.{~\ref{fig4}}(d) and (e), respectively]. For comparatively small detunings as in Fig.{~\ref{fig4}}(d), $I^-$ exhibits, at an injected power of about 0.51~mW, a SSB transition from a single to two asymmetric coexisting steady states [see panel (c)]. For comparatively large detunings as in Fig.{~\ref{fig4}}(e), on the other hand, $I^+$ exhibits two maxima in the range of injected power from 0.26~mW to 0.35~mW, which is evidence for a region of bistability with two coexisting symmetric steady states [see panel (b)].

Figures{~\ref{fig4}}(f,g) show corresponding plots obtained by simulating Eqs.~\eqref{eq:BHM}  while varying the normalized driving amplitude $f = 4F \sqrt{|U|}/\gamma^{3/2}$. We also set $U=\gamma/100$, which we estimated by comparing numerical and experimental standard deviations of the intensity around the SSB transition. Further rescaling the intracavity field as $(A,B) =(-2i \alpha^*_1\sqrt{|U|/\gamma},-2i\alpha^*_2\sqrt{|U|/\gamma})$ \cite{giraldo_the_2020} allows us to interpret $\gamma/|U|\sim100$ as the characteristic intracavity photon number; note that the thermodynamic limit corresponds to $\gamma/|U|\rightarrow \infty$, for which the mean field solutions---i.e., these without stochastic fluctuations---are strictly justified. Analytical results of bifurcation points of the mean field BHD derived in \cite{giraldo_the_2020} give an important handle to analyse the stochastic simulation results. Firstly, the mean field pitchfork bifurcation associated with the SSB transition occurs at $f \approx 4.04$ in panel~(f); this is consistent with the stochastic numerical simulations, which also indicate the presence of random switching between asymmetric states from this value up to $f \approx 5.5$. Secondly, the two predicted mean-field bifurcations of the symmetric state for $1.67 \lesssim f \lesssim 2.15$ clearly set the limits of the observed hysteresis cycle in Figure{~\ref{fig4}}(g).

\begin{figure}[t]
  \includegraphics[width=1 \columnwidth]{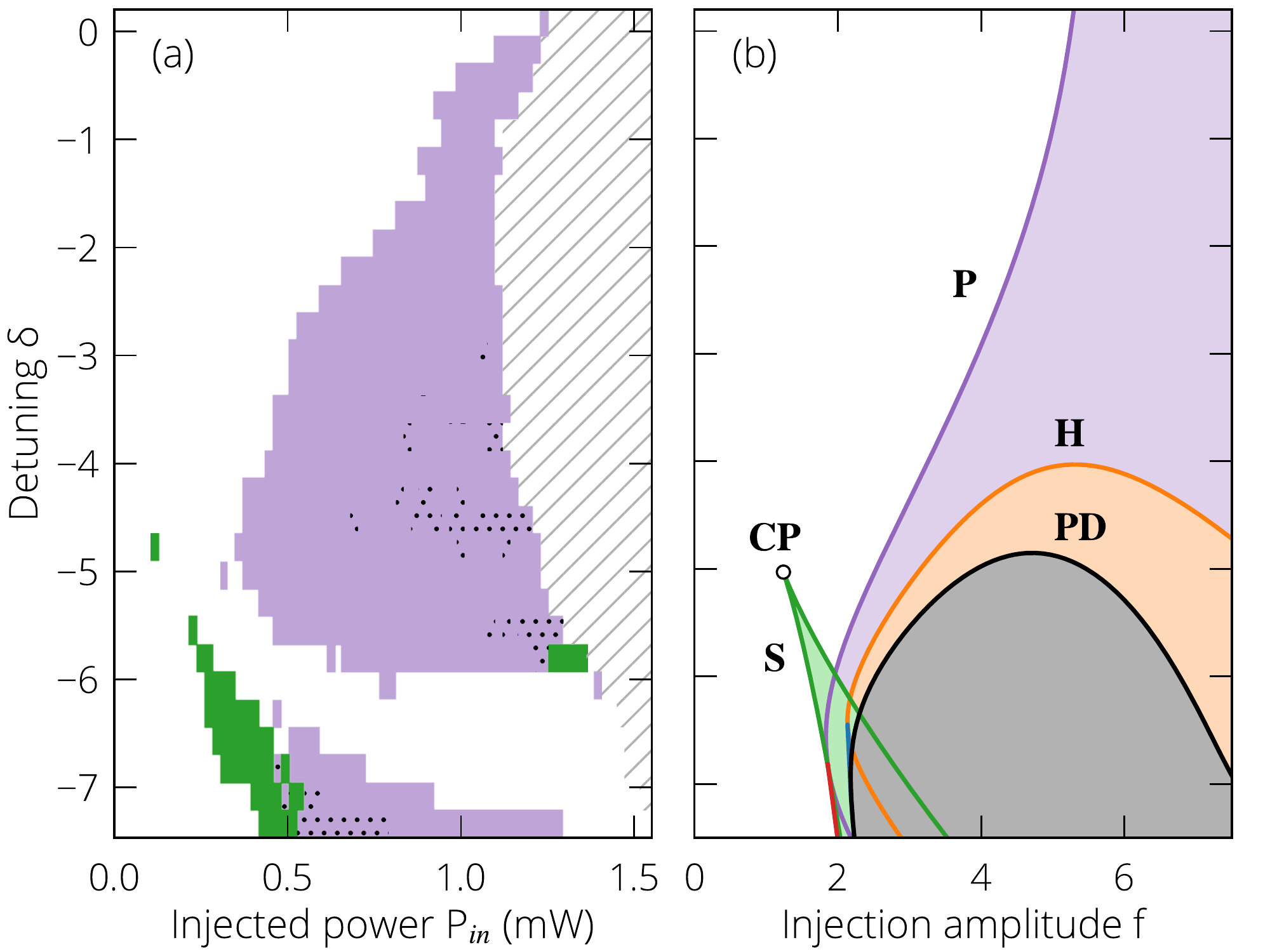}
  \caption{Two-parameter bifurcation diagrams in injection strength and detuning. (a) Experimental; colors stand for the number of maxima observed: white, one maximum for both $I^+$ and $I^-$; purple, two maxima for $I^-$; green, two maxima for $I^+$; dotted area, superposition of green and purple; hatched area, not investigated. (b) Numerical; curves represent different bifurcations: purple, pitchfork; green, saddle-node of symmetric steady states; orange, Hopf; black, period-doubling; shades of purple, regions with asymmetric solutions.}
  \label{fig5}
\end{figure}

We now identify in Fig.{~\ref{fig5}} the different regimes of operation in the two-dimensional parameter plane of driving strength and detuning. The experimental bifurcation diagram in panel~(a) was created by color coding the number of maxima found for $I^+$ and $I^-$. It reveals large regions of the parameter plane for which the system exhibits two asymmetric steady states; the SSB transition that bounds this region to the left moves towards lower injection power $P_{in}$ as the detuning $\delta$ is decreased; it is characterized by distinct maxima of $I^-$. Notice also the region of symmetric bistability (identified by maxima of $I^+$ only) that exists for smaller $\delta$; it may also overlap (dotted area) with asymmetric behavior: three states are observed (one lower symmetric and two higher asymmetric). We attribute the absence of asymmetric states for $-7 \lesssim \delta \lesssim -6$ to large deviations of the sample's position with regard to the laser beam at larger detunings. 

Figure~\ref{fig5}(b) shows the associated two-parameter bifurcation diagram of Eqs.~\eqref{eq:BHM} without noise (see Supplementary Material) computed using \textsc{Auto07p}~\cite{doedel_auto-07p_2007}. Overall, it is in good agreement with the experimental bifurcation diagram in terms of where symmetry-broken dynamics and bistability between symmetric states can be found. More specifically, the saddle-node bifurcation curve $\mathbf{S}$ with the cusp point $\mathbf{CP}$ at $(f,\delta) \approx (1.24,-5.03)$ delimits the region with bistability between symmetric steady states. The SSB transition is identified as a pitchfork bifurcation $\mathbf{P}$~\cite{cao_two_2016,casteels_quantum_2017}, yielding a region with two coexisting asymmetric steady states. As the detuning is decreased, the Hopf bifurcation curve $\mathbf{H}$ marks the onset of asymmetric oscillations, whereas $\mathbf{PD}$ indicates the first bifurcation of a period-doubling route to chaos, which eventually creates a pair of asymmetric chaotic attractors that are each confined to one cavities \cite{giraldo_the_2020}; they eventually collide, forming a symmetric chaotic attractor that features irregular switching between the two cavities \cite{giraldo_chaotic_2021}.


In conclusion, we have reported the first experimental realization of SSB in a coherently driven photonic dimer. For repulsive nonlinearities ($U>0$), we have shown that pure supercritical pitchfork bifurcations only exist for negative photon hopping energies ($J<0$); we also predict this to hold for $J>0$ provided that $U<0$. The SSB transition consists of the splitting of the two intracavity intensities under symmetric driving conditions; this is observed as two different randomly-selected states where light is localized in either of the two cavities. We mapped out the SSB transition in a large parameter plane and identified the region of bistability between symmetric states. Our experiments are in good agreements with the mean field Bose-Hubbard dimer model and stochastic simulations validated the random nature of this process. The theory for the BHD also predicts the existence of limit cycle oscillations and deterministic chaos. Our results constitute an important step towards the study of symmetry breaking in the few photon regime~\cite{casteels_quantum_2017}, at the crossroad between nonlinear dynamics and quantum optics.

B. G. and A. M. Y. would like to acknowledge the financial support from the European Union in the form of Marie Sk\l odowska-Curie Action grant MSCA-841351. This work is part of the research programme of the Netherlands Organisation for Scientific Research (NWO).  S.R.K.R. acknowledges an ERC Starting Grant with project number 85269.

\putbib
\end{bibunit}

\begin{bibunit}
\pagebreak
\onecolumngrid
\vspace*{0.5cm}
\begin{center}
\textbf{\large Supplementary Materials: Spontaneous symmetry breaking in a coherently driven nanophotonic Bose-Hubbard dimer}
\end{center}
\setcounter{equation}{0}
\setcounter{figure}{0}
\setcounter{table}{0}
\setcounter{page}{1}
\makeatletter
\renewcommand{\theequation}{S\arabic{equation}}
\renewcommand{\thefigure}{S\arabic{figure}}
\renewcommand{\bibnumfmt}[1]{[S#1]}
\renewcommand{\citenumfont}[1]{S#1}

\section{Numerical observation of SSB in the presence of noise}

\begin{figure}[h]
  \centerline{\includegraphics[width=0.5 \columnwidth]{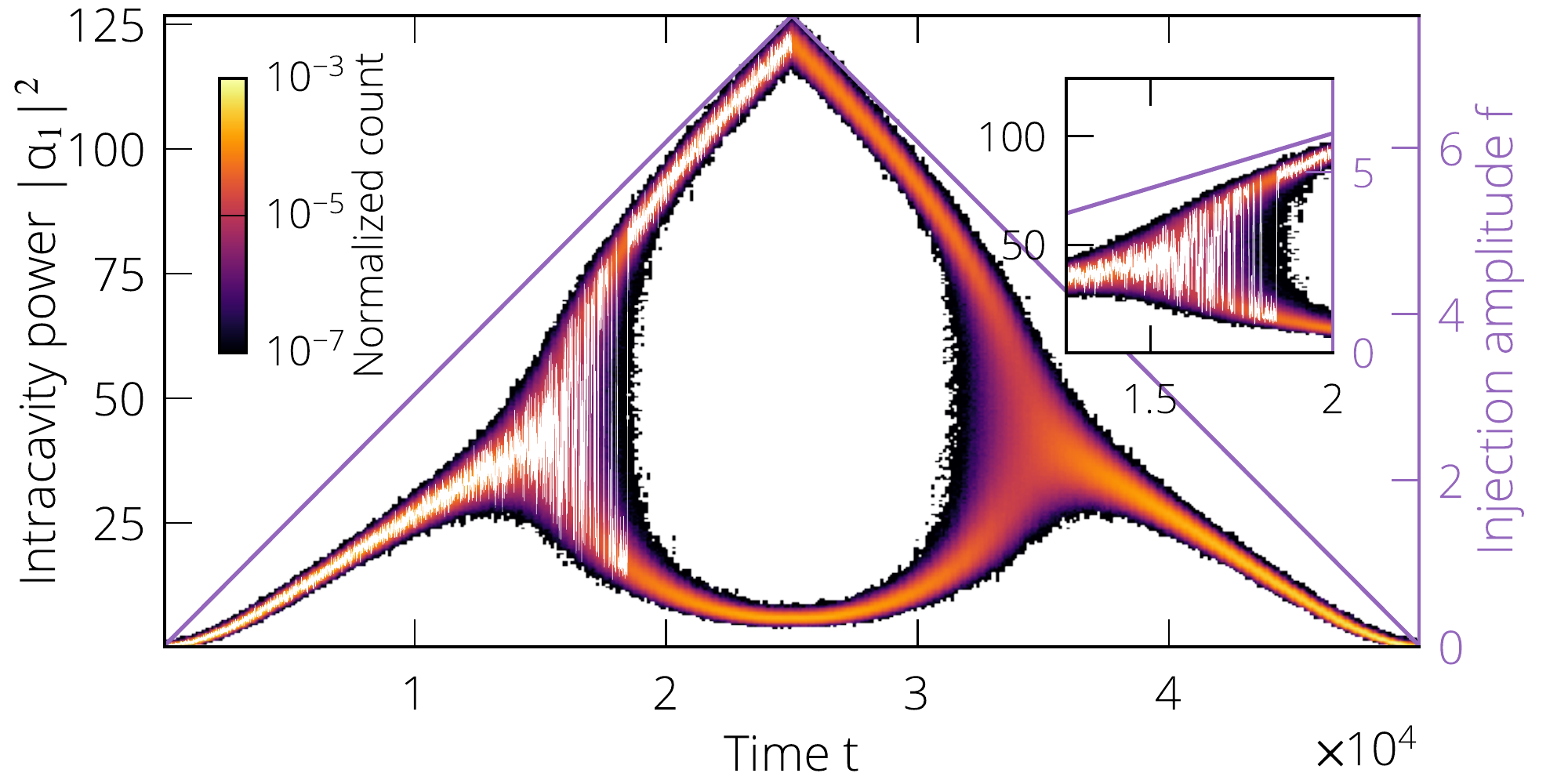}}
  \caption{Temporal observation of SSB in stochastic simulations of Eqs.~(1), showing normalized histograms of the output power of a single cavity, computed over 500 realizations, with modified driving power (purple curve). The white curve shows a single realization and highlights switches that occur in the vicinity of the SSB transition; see also the inset panel. $\delta=-1.59$, $\kappa=3.3$, $\gamma=1$, $U=\gamma/100$, $D = \sqrt{\gamma/2}$.}
  \label{fig:SSBtrace}
\end{figure}

To reproduce the experimental observations of SSB, we perform stochastic simulations with the xSPDE \textsc{Matlab} toolbox~\cite{xSPDE} by integrating Eqs.~(1) with a fourth-order Runge-Kutta method, where the time steps are of size $\gamma^{-1}/10$. In all our stochastic simulations we set $D=\sqrt{\gamma/2}$ to account for quantum fluctuations in the truncated Wigner approximation~\cite{carusotto_quantum_2013}. The value of $U$ was then estimated by matching $\mathrm{std}\left(|\alpha_i|^2\right)/\mathrm{mean}\left(|\alpha_i|^2\right)$ near the SSB transition to the experiment. A $550$~MHz filter was applied to match the experimental detector bandwidth. Figure~{\ref{fig:SSBtrace}} shows a histogram associated with a single cavity built from 500 computed realizations of Eqs.~(1) as the modified driving power is ramped up and down; compare with Fig.~3. In the single shown trajectory the noise $\zeta_{1,2}(t)$ is seen to induce fast random switching between the two states near the the SSB transition. Consequently, the final state of the cavities is selected at random after the SSB transition occurs  at $t\approx$~1.5x10$^4$. Note that for a symmetric system such as Eqs.~(1), the two asymmetric states are observed with equal probability and the histograms associated with each of the two cavities are equivalent. 

\section{Identification of bistability of symmetric and asymmetric states}

\begin{figure}[ht]
  \centerline{\includegraphics[width=0.5 \columnwidth]{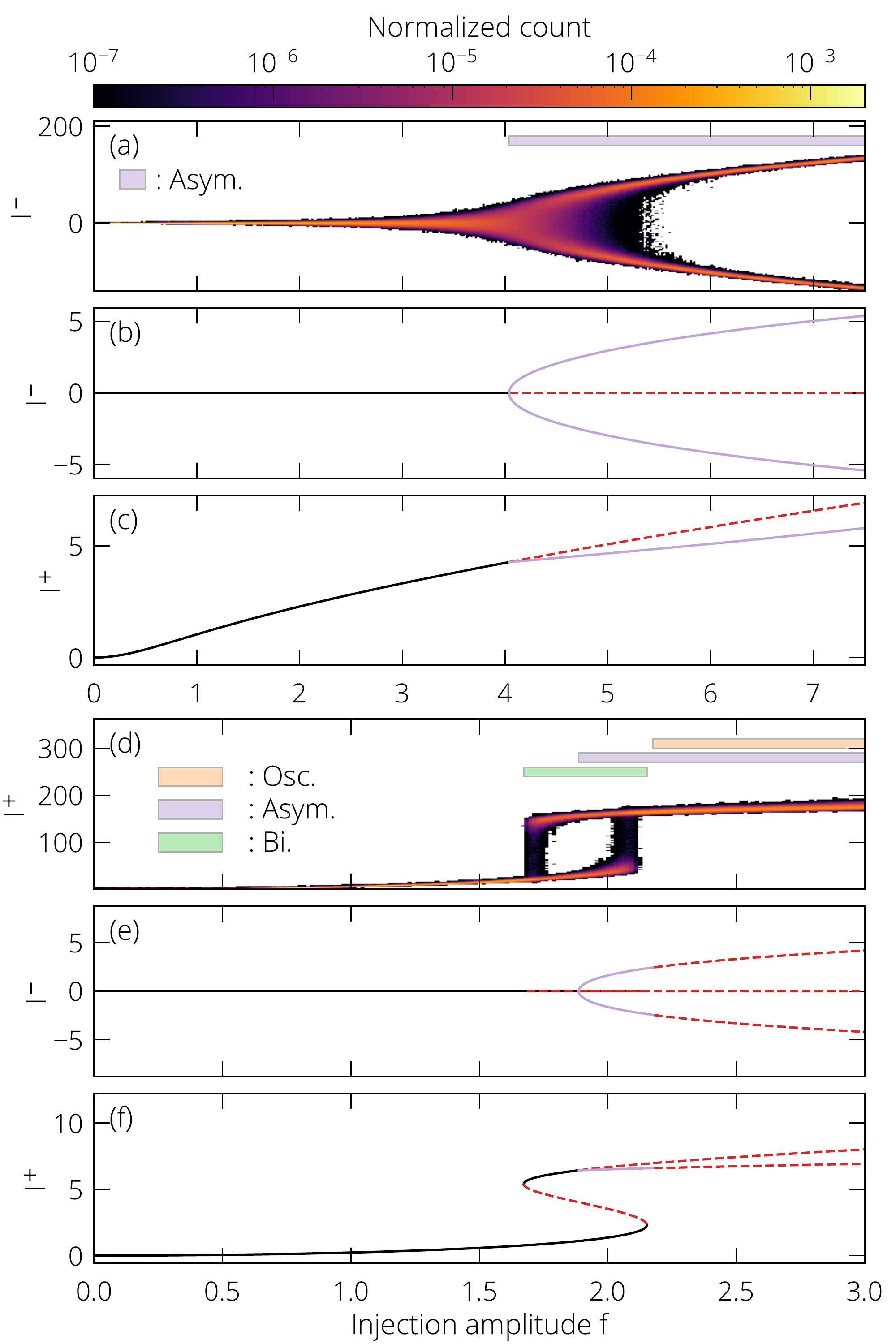}}
  \caption{(a),(d) Histogram over $f$ computed from 500 realizations of Eqs.~(1) showing SSB as represented by $I^-$ and hysteresis of symmetric steady states as represented by $I^+$, respectively. Panels (b),(c) and panels (e),(f) show the associated steady states computed with the numerical continuation package \textsc{Auto07P} in terms of both $I^-$ and $I^+$. (a)-(c) $\delta=-2.86$; (d)-(f) $\delta=-6.19$. }
  \label{fig:Iminusplus}
\end{figure}

We now demonstrate with Figure~\ref{fig:Iminusplus} that the difference $I^-$ and the sum $I^+$ are suitable observables to identify and distinguish, respectively, SSB with asymmetric steady states and hysteresis between symmetric steady states; compare with Fig.~4 and Fig.~5.  More specifically, Fig.~\ref{fig:Iminusplus} shows in panels~(a) and~(d) histograms over the injection amplitude $f$, computed from 500 realizations of Eqs.~(1), for both detunings used in Fig.~4. The case of SSB for $\delta=-2.86$ is best identified by the difference $I^-$, as can be seen clearly by comparing in Fig.~\ref{fig:Iminusplus}(a) with the computed branches of associated steady states in panels~(b) and~(c); in particular, the sum $I^+$ does not show SSB clearly. In contrast, the range of bistability between symmetric steady states for $\delta=-6.19$ is best identified by $I^+$; compare Fig.~\ref{fig:Iminusplus}(d) with the computed branches of associated steady states in panels~(e) and~(f) and note that $I^-$ does not distinguish this bistability at all. These observations are further strengthened by the fact that the unstable states (dashed red curves) are not observed in experiments.

\section{Normalized model without noise}
Equations~(1) of our main manuscript may be further normalized to~\cite{giraldo_the_2020}:

\setcounter{equation}{1}

\begin{equation}
\frac{\partial E_{1,2}(t)}{\partial t} = \left[-1 + i\left(\delta + \sgn(U) |E_{1,2}|^2\right)\right] E_{1,2}+ \quad+ i\kappa E_{2,1} + f
\label{eq:main}
\end{equation}
for the evolution of the slowly varying intracavity electric field envelopes $E_{1,2}$ of each of the two cavities in a frame rotating at $\omega_p$. Here, the time $t$ is normalized to the cavity dissipation rate $\gamma$, and we defined the normalized detuning, coupling and driving field as $\delta = -2(\omega_p - \omega_c)/\gamma$, $\kappa=-2J/\gamma$, $f = 4F \sqrt{|U|}/\gamma^{3/2}$, respectively; here we assume $F= F_{1}=F_{2}$. Equation~\eqref{eq:main} is related to the BHD model Eqs.~(1) via the transformation:
 \begin{equation}
 \alpha_{1,2} \mapsto E_{1,2} = -2i\alpha_{1,2}^* \sqrt{|U|/\gamma}.
 \end{equation}
We remark that this model is reminiscent of the celebrated Lugiato-Lefever equation for passive optical resonators~\cite{lugiato_spatial_1987}.

\section{Observation of pure SSB: positive versus negative coupling}

\begin{figure}
  \centerline{\includegraphics[width=1 \columnwidth]{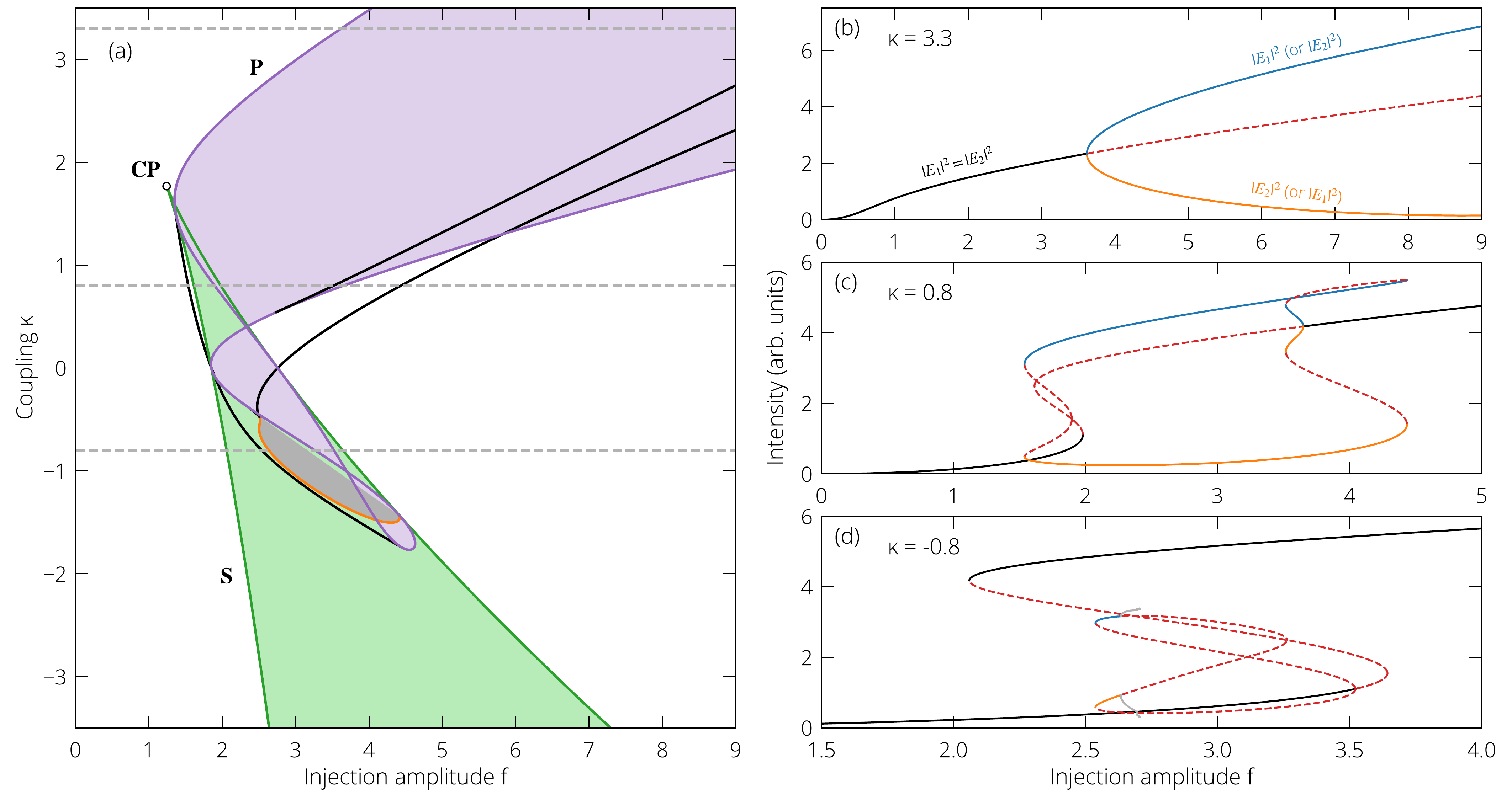}}
  \caption{(a) Two-parameter bifurcation diagram in the $(f,\kappa)$-plane of Eqs.~(1) for $\delta=-3.5$ with regions of bistability of asymmetric steady states (purple shading) and of symmetric steady states (green shading). (b),(c),(d) One-parameter bifurcation diagrams in $f$ for $\kappa$ as indicated by the grey dashed lines in panel (a).}
  \label{fig:2pardiag}
\end{figure}

Figure~\ref{fig:2pardiag} presents evidence for the statement that SSB is best observed for coupling $\kappa>0$ and sufficiently large. Indeed, the two-parameter bifurcation diagram in the $(f,\kappa)$-plane of Eqs.~(1) in Fig.~\ref{fig:2pardiag}(a) shows that, in that case, one does not find any bistability between symmetric steady states. As a result, the one-parameter bifurcation diagram for $\kappa=3.3$ in panel~(b) shows symmetry breaking at the pitchfork bifurcation very clearly, and there are no other bifurcation in this slice. Similarly, for sufficiently negative $\kappa$, one only observes bistability between two symmetric steady states. However, Fig.~\ref{fig:2pardiag}(a)  shows also that there is a range of $\kappa$ near $0$ where one finds both SSB and hysteresis between symmetric steady states. Examples of one-parameter bifurcation diagrams that feature both are shown in panel~(b) for $\kappa=0.8$ and in panel~(d) for $\kappa=- 0.8$, respectively. 

We remark that this ``mixing'' between the two types of bistability is considerably enhanced in Fig.~\ref{fig:2pardiag}(a) because we consider here the smaller value of $\delta=-3.5$, compared to the value of $\delta=0$ of Fig.~1(c). Nevertheless, the case with a ``pure" (pitchfork bifurcation) SSB transition is still observed at larger $\kappa$ values; moreover, for smaller values of $\kappa>0$ the states born from the SSB still exist and, while they now coexist with a bistability of symmetric states, the difference between their intensities is still large [see blue and orange curves in panel (c)]. Finally, the region of symmetry broken states extends over a small region for $\kappa<0$, where it is strictly inside the region of coexisting symmetric steady states. We remark that it is generally subject to coexistence with other states (e.g., oscillating orbits and chaos) \cite{giraldo_the_2020}, which implies that one cannot observe a ``pure'' pitchfork bifurcation of SSB for $\kappa<0$.

\putbib
\end{bibunit}


\end{document}